# SUPERFLUID HELIUM TESTING OF A STAINLESS STEEL TO TITANIUM PIPING TRANSITION JOINT

W. Soyars[1], A. Basti[3], F. Bedeschi[3], J. Budagov[2], M. Foley[1], E. Harms[1], A. Klebaner[1], S. Nagaitsev[1], and B. Sabirov[2]

[1]Fermi National Accelerator Laboratory
Batavia, IL, 60510, USA

[2]Joint Institute for Nuclear Research
Dubna, 141980, Russia

[3]National Institute of Nuclear Physics
Pisa, 56127, Italy

**ABSTRACT**

Stainless steel-to-titanium bimetallic transitions have been fabricated with an explosively bonded joint. This novel joining technique was conducted by the Russian Federal Nuclear Center, working under contract for the Joint Institute for Nuclear Research. These bimetallic transitions are being considered for use in future superconducting radio-frequency cavity cryomodule assemblies. This application requires cryogenic testing to demonstrate that this transition joint remains leak-tight when sealing superfluid helium. To simulate a titanium cavity vessel connection to a stainless steel service pipe, bimetallic transition joints were paired together to fabricate piping assemblies. These piping assemblies were then tested in superfluid helium conditions at Fermi National Accelerator Laboratory test facilities. The transition joint test program will be described. Fabrication experience and test results will be presented.

## INTRODUCTION

International research and development to support the International Linear Collider is underway. The issue considered here is the joining of dissimilar metals during Superconducting Radio Frequency (SRF) cryomodule construction, specifically the feasibility of techniques that would allow a titanium (Ti) SRF vessel to join to a stainless steel (SS) supply tube. Since SS and Ti cannot be joined by conventional electron-beam methods, alternative methods are under investigation and in use [1]. This has been successfully done for other SRF accelerator components [2]. A joint program involving JINR (Dubna, Russia), FNAL (Batavia, United States), INFN (Pisa, Italy), and IEP (Sarov, Russia) was conducted for further study [3]. The focus was joining coaxial tubes, not flat plates or sheets.

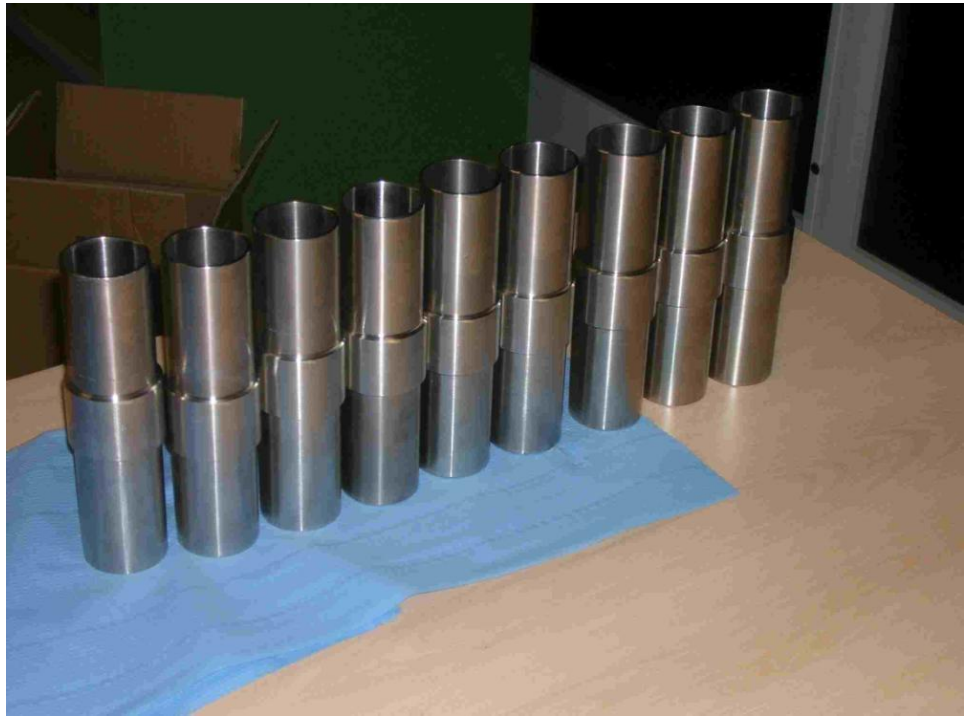

**FIGURE 1.** SS-to-Ti Transition Joint Tubes.

Stainless steel-to-titanium bimetallic transitions have been fabricated. A stainless steel sleeve covers the junction and is then explosion bonded onto the external surface of 316L SS and Ti Grade 2 tubes. See FIGURE 1. The design, fabrication, and characteristics of these bimetallic coaxial tube transitions are discussed elsewhere [4-7]. See TABLE 1 for description and geometry.

Project requirements have established the maximum integral leak rate for the transition tubes. It is not to exceed $1 \times 10^{-10}$ Pa-m$^3$/sec at room temperature and $1 \times 10^{-9}$ Pa-m$^3$/sec at LN$_2$ temperature. The goal is to test the transition tubes in superfluid helium conditions at Fermilab to see how well they seal under actual service conditions.

**TABLE 1.** Ti-SS bimetallic tubes for test program.

| Assembly Designation | Fabrication Year | Tube Designation | OD [mm] | Wall Thick. [mm] | Length [mm] |
|---|---|---|---|---|---|
| I | 2008 | 5A, 6A | 47 | 1.95 | 190 |
| II | 2008 | 8A, 9A | 47 | 1.95 | 190 |
| III | 2008 | 3N, 7N | 47 | 1.95 | 190 |
| IV | 2009 | 5, 6 | 60 | 2.5 | 170 |
| V | 2009 | 9, 11 | 60 | 2.5 | 170 |

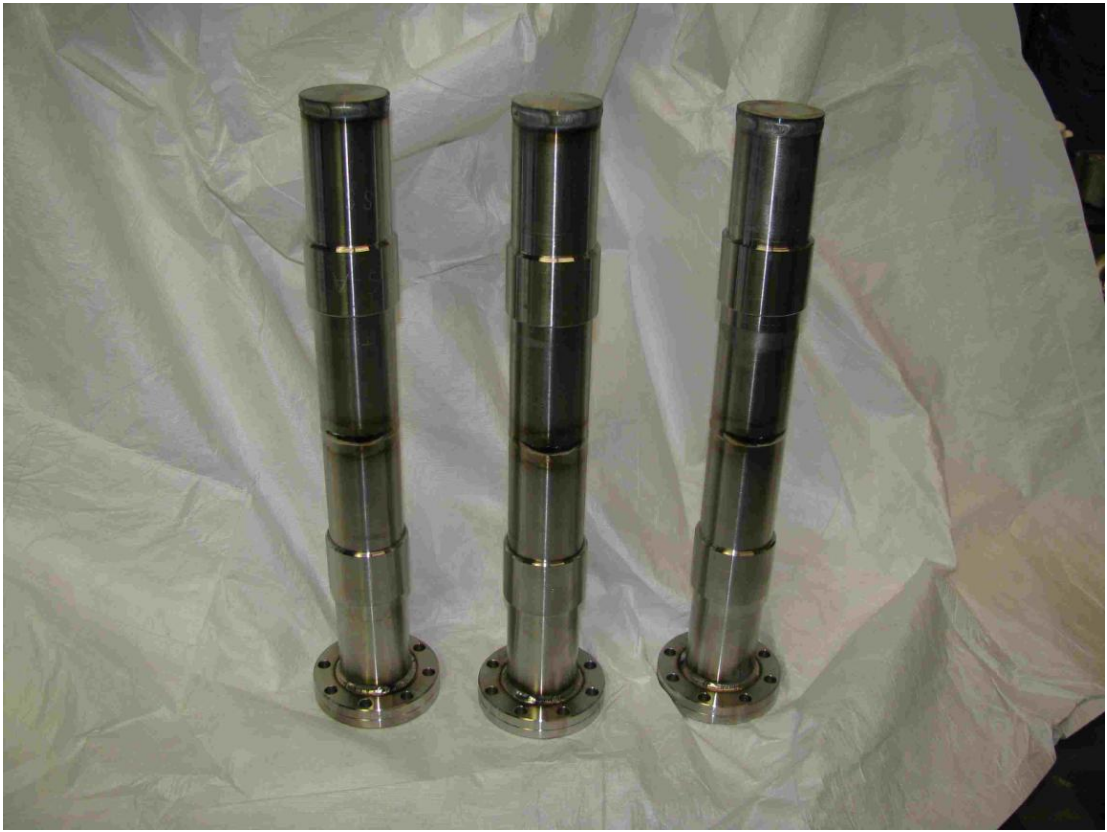

**FIGURE 2**. The Transition Joint Test Assemblies, 47 mm OD version. Each has a pair of transition joints.

## ASSEMBLY FABRICATION EXPERIENCE

The final assembly of test articles was completed at Fermilab. Five bimetallic Transition Joint Test Assemblies were made for attachment to the cryogenic test system, each with a pair of transition tubes. See FIGURE 2. Also noted in TABLE 1, two versions of this size Transition Joint are available: non-annealed or annealed (after the explosion-bonding procedure).

### Initial Leak Checking

Before a bimetallic transition tube was shipped to Fermilab for a final assembly, it was qualified as leak-tight at various conditions and multiple thermal cycles at INFN/Pisa and at JINR/Dubna [8, 9]. These successful initial results were repeated at Fermilab. See TABLE 2 for initial values.

Note, some transition samples were rejected because they did not pass the initial leak check after the joining process. For the tubes made in 2009, five out of thirteen total passed. It is believed that better tolerancing of the Ti wall thickness may help improve this acceptance percentage [9].

### Welding Temperature

Transition Joint Test Article Assembly titanium tubes were welded in an argon atmosphere. Even though no specification on a temperature limit for the explosion-bond exists, means to minimize the temperatures on the stainless steel collars during welding were investigated. Water-cooled copper heat sinks were fabricated. When cooling water was used, peak temperature was < 323 K (50 °C). When welded without water cooling, peak temperature was < 343 K (70 °C). The mild temperature increase with gas cooling only suggests that these bimetallic tubes can be welded in real conditions without any special cooling, without any risk. The use of active cooling had been important to the success of other bimetallic transition usage in cryomodules. [2]

**Post-fabrication Ambient Leak Checks: Effect of Ultrasonic Cleaning**

Since the Transition Joint Test Assembly will be installed for testing on a vacuum system that normally is used for SRF cavity beam tubes, they must be particulate free. They received ultrasonic water cleaning. Note, none of the 47 mm transition tubes available in 2008 had previously been cleaned ultrasonically. Post-fabrication and cleaning leak check results are shown in TABLE 2. In both cases in which a leak was found, the leak was at the Ti end of the explosion joint on transition tube.

For *Tube 8A* and *7N*, the hypothesis is that the bimetallic explosion-joint may have been damaged during the ultrasonic cleaning. The concern is that, at the explosion-joint, one can destroy the structure of the diffusion layer, which is only 50-100 microns thick. The leak on tube *7N* actually meets the room temperature leak rate requirement. However, a positive leak indication is seen.

Based on this experience, some 60 mm transition tubes shipped in 2009 underwent extensive ultrasonic cleaning at Pisa before shipment to Fermilab. Three total samples (*9, 11* and *18*) were ultrasonically cleaned. Two (*9, 11*) were cleaned in cold distilled water-- the normal procedure--and were proven to have their leak-tightness unaffected. The third sample (*18*), was aggressively cleaned with 100 C water and afterwards showed a slight leak detected at ~$5 \times 10^{-11}$ Pa-m$^3$/sec sensitivity, similar to what was experienced with tube *7N* [8]. However, the drastic change in leak rate, as observed for *Tube 8A*, was not repeated. The 60 mm versions of the Transition Joint Test Assemblies were ultrasonically cleaned by Fermilab in preparation of installation and cooldown. Afterwards, both assemblies (IV and V with four tubes) proved to be leak-tight.

There have been no changes to the explosion joint design to specifically address this issue. However, future joints will be tested for this condition and design tolerances to improve the fabrication acceptance rate may help.

**TABLE 2.** Leak check results after ultrasonic cleaning.

| Assembly Designation | He leak rate at Ambient & 77 K, Initial [Pa-m$^3$/sec] | He leak rate at Ambient Temp <u>after Ultrasonic Clean</u> [Pa-m$^3$/sec] | Comment | Tube with Leak |
|---|---|---|---|---|
| I | < $5 \times 10^{-11}$ | < $5 \times 10^{-11}$ | OK, no leak | None |
| II | < $5 \times 10^{-11}$ | ***~$1 \times 10^{-7}$*** | ***Leak*** | 8A |
| III | < $5 \times 10^{-11}$ | ***~$5 \times 10^{-11}$*** | ***Slight leak*** | 7N |
| IV | < $3 \times 10^{-11}$ | < $3 \times 10^{-13}$ | OK, no leak | None |
| V | < $3 \times 10^{-11}$ | < $4 \times 10^{-13}$ | OK, no leak | None |

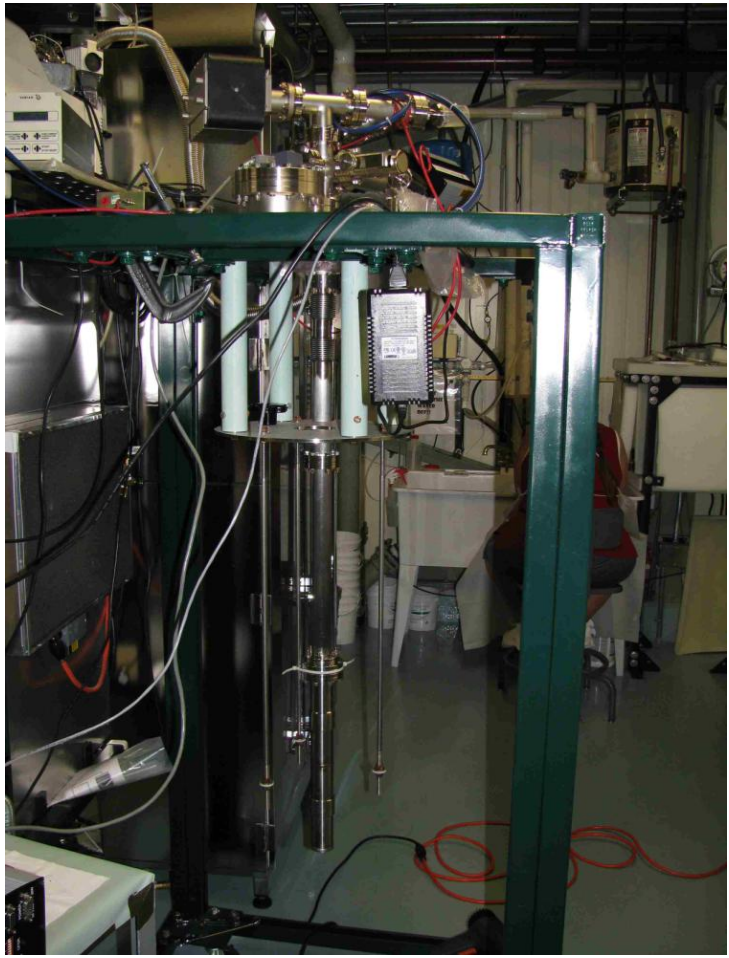

**FIGURE 3**.  Transition Joint Test Assembly, prior to installation into the A0 Vertical Test Dewar.

## SUPERFLUID TEST FACILITIES FOR ASSEMBLY LEAK CHECKS

Two different test beds are available to cool a SS-to-Ti Transition Joints assembly to superfluid temperatures:  the A0 Vertical Test Dewar (A0 VTD) and the Meson Detector Building Horizontal Test System (MDB HTS) [10, 11]. The general test plan is to confirm at room temperature that the assembly is initially leak free, then continuously monitor the leak-rate during cooldown and superfluid operation.

The primary test bed is the A0 VTD.  See FIGURE 3.  The interior of the Transition Joint can be connected to a high vacuum Residual Gas Analyzer for a direct leak rate measurement while the exterior is cooled down and bathed in superfluid.  At the secondary test bed, the MDB HTS, the transitions are tested in actual service conditions with superfluid helium on the interior while surrounded by insulating vacuum. However, the observed leak rate assessment at MDB HTS is not as straightforward since these checks must be made on the entire cryomodule system's insulating vacuum, something for which it was not designed.

## SUPERFLUID TEST RESULTS

The Transition Joint Test Assemblies have been cooled down with superfluid helium on various occasions. There was continuous monitoring for leaks. Overall results are given in TABLE 3.

**TABLE 3.** Bimetallic Transition Joint Assembly leak check results, at superfluid conditions.

| Assy | Test System | Initial He leak rate, ambient [Pa-m$^3$/sec] | Duration of 2K conditions [hr] | He leak rate, superfluid [Pa-m$^3$/sec] | Final He leak rate, ambient [Pa-m$^3$/sec] | Conclusion |
|---|---|---|---|---|---|---|
| I | A0 VTD | $< 5 \times 10^{-10}$ | 0.3 | $< 5 \times 10^{-10}$ | $< 2 \times 10^{-11}$ | No leak observed. |
| I | MDB HTS | $< 7 \times 10^{-11}$ | 30 | $< 3 \times 10^{-11}$ | $< 3 \times 10^{-11}$ | No leak observed. |
| I | A0 VTD | $< 6 \times 10^{-13}$ | 1.5 | $< 6 \times 10^{-13}$ | $< 10^{-12}$ | No leak observed |
| I | A0 VTD | $< 2 \times 10^{-11}$ | 21 | **$1 \times 10^{-8}$** | **$1 \times 10^{-12}$** | **Cold leak, and subsequent room temp. leak** |
| V | A0 VTD | $< 5 \times 10^{-13}$ | 23 | $< 3 \times 10^{-11}$ | $< 10^{-12}$ | No leak observed |
| IV | A0 VTD | $< 7 \times 10^{-13}$ | 20 | $< 8 \times 10^{-11}$ | ----- | No leak observed |

The initial test indicated no leak, yet it was performed in the presence of some new Vertical Test Dewar hardware, which subsequently was baked-out to improve sensitivity. The second test, at MDB HTS, was successful and showed no leak for many hours. This test was terminated when a suspected cryomodule component cold leak (into the common insulating vacuum and thus leak detector) occurred during the second day of the test; this flange leak was identified and repaired during a subsequent warmup. The third test went well with no observed leaks, although it was of limited duration. The fourth test, shown in FIGURE 4, did reveal a cold leak. The RGA response coincided with the supply dewar running empty; perhaps some warming of the vapor space pumping line would occur as flow ceases, and then some accumulated He trapped on a cold surface was then released for detection. Furthermore, the subsequent ambient temperature leak-check after this warmup confirmed the presence of a small leak. The final two tests were conducted with the most current transition design. These tests at superfluid conditions were successful with no leak indicated.

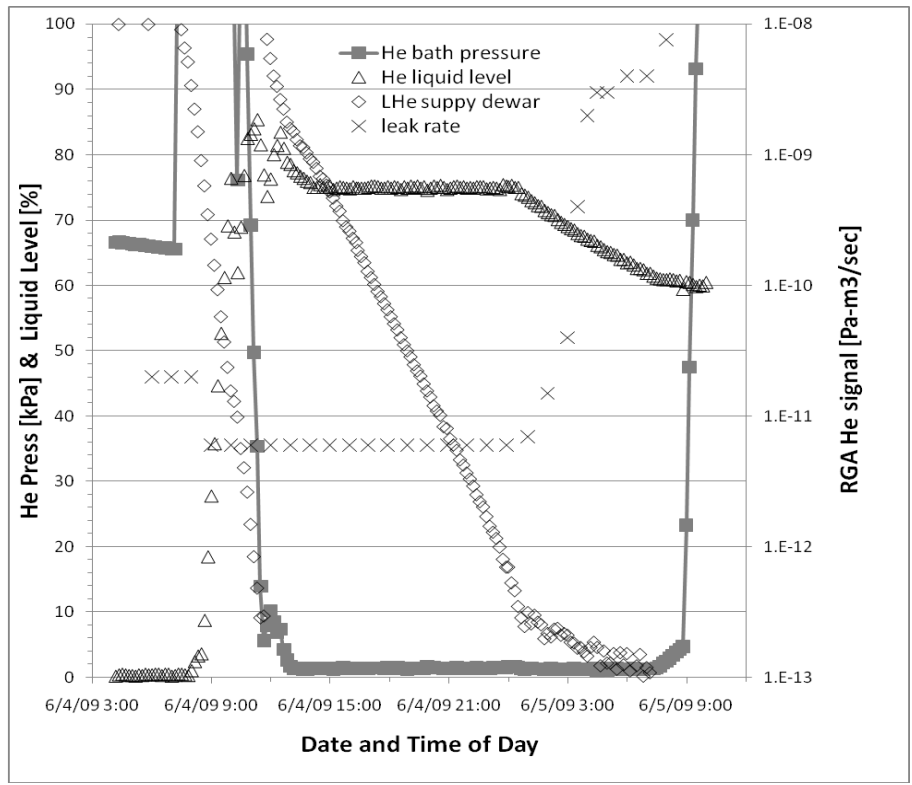

**FIGURE 4.** The SS-Ti Transition Test Assembly at superfluid conditions. Fourth test, indicating leak.

## CONCLUSIONS

JINR-Dubna delivered bimetallic Ti-SS tubes to Fermilab to assess their ability to contain superfluid helium. Seven 47 mm tubes and five 60 mm tubes were delivered as test articles. These bimetallic tubes were verified as being initially leak tight.

The tubes were fabricated into five assemblies by having the tubes welded in pairs for installation into a cryogenic test system. Three assemblies (six tubes) were fabricated with no change in its leak rate. However, two assemblies, of the earlier generation, had ambient condition leaks develop at a bimetallic joint, which was attributed to an ultrasonic cleaning process. For future fabrication, design tolerances to improve the fabrication acceptance rate may help.

On several occasions, a pair of tubes was cooled down and operated in a superfluid environment. The explosion-bonded SS-to-Ti transitions joints successfully contained superfluid. However, one long duration test on the original transition design configuration indicated a leak. Superfluid leak testing with four of the most recently fabricated transition tubes successfully indicated no leak. These results will contribute to further development. This demonstrates that the design could be feasible for superfluid helium service, but more testing and development is required.

While long duration testing was limited during this program due to test bed schedule requirements, we see the importance of such testing. So a future test program should emphasize longer term tests. Also, these results suggest that several thermal cycles with superfluid are valuable.

This program has been working with Ti-to-SS transitions. Future SRF cryomodule designs are considering the use of Niobium-to-SS transitions. The cryogenic testing experience gained here will be useful for future superfluid testing of this concept. Facilities and methods developed here will be useful to support a test program for future transition piece assessments.

## ACKNOWLEDGEMENTS

One should acknowledge the technical groups who contributed to making these tests possible. At the Russian Federal Nuclear Center (Sarvov), the technical staff manufactured these unique samples of bimetallic Ti to SS joints. At Fermilab, the Accelerator Division Cryogenic and A0 Mechanical Support SRF groups fabricated support hardware for the transitions, welded the pairs together, completed assemblies, performed initial leak checking, and provided cold test support.

Fermilab is operated by Fermi Research Alliance, LLC under Contract No. DE-AC02-07CH11359 with the United States Department of Energy.